\documentclass[12pt, preprint]{aastex}
\usepackage{natbib}

\def\be{\begin{equation}}
\def\ee{\end{equation}}

\begin{document}

\shorttitle{On the pulse-width ... } \shortauthors{Gil}

\title{On the pulse-width statistics in radio pulsars}

\author{Marcin Kolonko\altaffilmark{1,2}, Janusz Gil\altaffilmark{3} and Krzysztof Maciesiak\altaffilmark{3}}
\altaffiltext{1}{Institute of
Nuclear Physics, Radzikowskiego 152, 31-342 Krak\'ow,
Poland}\altaffiltext{2}{Institute of Astronomy, Jagiellonian
University, Orla 171, 30-244 Krak\'ow, Poland}
\altaffiltext{3}{Institute of Astronomy, University of Zielona
G\'ora , Lubuska 2, 65-265 Zielona G\'ora, Poland}

\baselineskip=25pt
\begin{abstract}

The Monte Carlo simulations of pulsar periods, pulse-widths and
magnetic inclination angles are performed. Using the available
observational data sets we study a possible trial parent
distribution functions by means of the Kolmogorov-Smirnov
significance tests. We also use an additional condition that the
numbers of generated interpulses, whether from both magnetic poles
or from single pole, are at the observed levels. We conclude that
the parent distribution function of magnetic inclination angles is
neither flat nor cosine but it is a more complicated function with
a local maximum near $\alpha=25^\circ$ and another weaker one near
$\alpha=90^\circ$. The plausible distribution function of pulsar
periods is represented by the gamma function. The beaming fraction
describing the fraction of observable radio pulsars is about 0.12.

\end{abstract}

\section{Introduction}

Statistical studies of the pulse-width in mean profiles of radio
pulsars are an important tool for investigations of the pulsar
radiation geometry. One especially important parameter that can be
derived from such studies is the inclination angle between the
magnetic and the spin pulsar axes. Early studies were carried out
by \citet{hp69,rs72,rs73,b76} and \citet{ml77}. Since the amount
of the available data was small, these papers suffered from
problems of small number statistics. A more complete work was
performed by \citet{p79} and \citet{lm88}, who analyzed samples of
about 200 pulse-width data measured near 400 MHz. Although the
database used in these papers was quite rich, the pulse-width
measurements were contaminated by the interstellar scattering
dominating at low radio frequencies. More recently \citet[][ GH96
hereafter]{gh96} compiled a new database of 242 pulse-widths
$W_{10}$ (corresponding to about 10\% of the maximum intensity)
measured at a higher radio frequency (near 1.4 GHz), which was
relatively unbiased as compared with the lower frequency data.
GH96 used their pulse-width database to perform Monte Carlo
simulations in an attempt to derive the distribution statistics of
pulsar periods, pulse widths, magnetic inclination angles and
rates of the interpulse occurrence. They concluded by comparing
the simulated and observed (or observationally derived) quantities
that the observed distribution of the inclination angles resembles
the sine function following from the flat (random) distribution in
the parent population, and that the probability (beaming fraction)
of observing a pulsar was about 0.16. GH96 also pointed out that
the rates of interpulse occurrence should be considered as an
important aspect of pulsar population studies.

On the other hand, \citet[][ henceforth TM98]{tm98} using a
different method based on an analysis of the indirectly derived
polarization position angles and magnetic inclination angles
concluded that the observed distribution of the latter is
cosine-like rather than the sine-like suggested by GH96. They also
obtained the beaming factor $0.1\pm 0.02$, considerably lower than
0.16 obtained by GH96. TM98 pointed out a likely source of this
discrepancy, namely the incorrect assumption used by GH96 that the
observed distribution and the parent distribution of pulsar
periods are similar. Recently, \citet[][ henceforth ZJM03]{zjm03}
followed the Monte-Carlo simulation scheme developed by GH96.
ZJM03 argued that both the parent distribution function and the
observed distribution of pulsar periods can be modelled by the
gamma function but with different values of free parameters, and
their Monte-Carlo simulations included searching for a 2-D grid of
these parameters. As a result, ZJM03 concluded that indeed the
cosine-like distribution (suggested by TM98) is a much more
suitable to model the inclination angles in the parent pulsar
population than the flat distribution (suggested by GH96). They
argued that a most plausible parent distribution is a modified
cosine function, which has a peak around $25^\circ$ and another
weaker peak near $90^\circ$. They also obtained the beaming factor
$\sim 0.12$, consistent with the result of TM98.

As emphasized by ZJM03 in the conclusions of their paper, neither
these authors nor TM98 considered potentially important
constraints related to the observable interpulse emission.
Although GH96 did consider the issue of interpulse emission, but
their estimate of rates of occurrence was not quite correct (see
\S 3.4 in this paper). Moreover, their statistical analysis was
biased by incorrect assumption mentioned above concerning parent
distribution of pulsar period. In this work we follow the
simulation scheme of ZJM03 but we include the actual rates of
interpulse occurrences. We demonstrate that a pure cosine
distribution of the magnetic inclination angles generates much too
few interpulses as compared with observations and should be
rejected as a plausible distribution function. We found that the
modified cosine function of ZJM03 (see eq.~[9]) is much better in
this respect, mainly because of the weak second maximum near
$\alpha=90^\circ$. Both the above distribution functions can
reproduce the observed distributions of pulse-widths, pulsar
periods and inclination angles almost equally well, and thus the
observed interpulse statistics provides the most restrictive
constraint discriminating between different trial distribution
functions.

\section{Basic formulae}

\subsection{Pulse-widths}

It is generally accepted that the pulsar radio emission is
relativistically beamed along the open dipolar field lines (to
within $1/\gamma$, where $\gamma\sim 100$ is the Lorentz factor of
the emitting sources). Thus, the pulse width $W_{10}$ at the level
of 10\% of the maximum profile intensity can be written as \be
W_{10}=4\arcsin\left\{\frac{\sin[(\rho_{10}+\beta)/2]\sin
[(\rho_{10}-\beta)/2]}{\sin\alpha\cdot\sin(\alpha+\beta)}\right\}^{1/2}
\label{w10} \ee \citep{g81}, where $\rho_{10}$ is the beam-width
corresponding to 10\% intensity level, $\alpha$ is the inclination
angle of the magnetic axis to the spin axis, $\beta$ is the impact
angle of the closest approach of the line-of-sight to the magnetic
axis, and thus $\xi=\alpha+\beta$ is the observer angle between
the spin axis and the line-of-sight. It is worth noting that
equation~(\ref{w10}) assumes the symmetry of a pulsar beam with
respect to the fiducial plane containing both the spin and the
magnetic pulsar axes. Thus, the appropriate pulse widths database
should contain only pulsars with symmetrical profiles for which
$W=2\varphi$, where $\varphi$ is the pulse longitude measured from
the fiducial phase corresponding to the fiducial plane (e.g.
GH96).

\subsection{Opening angles}

The opening angle of the radio beam (beam-width) corresponding to
10\% intensity level can be derived from pulse width measurements
$W_{10}$ (and $\alpha$, $\beta$ values derived from the
polarization data) in the form of the so-called $\rho-P$ relation.
\citet{lm88} obtained $\rho_{10}\approx 6.^\circ 5P^{-1/3}$ for
the pulse-width data at 408 MHz, which scaled to 1.4 GHz writes
\be \rho_{10}\approx 5.^\circ 8P^{-0.33} \label{ro10}. \ee
However, \citet{b90} reanalyzed the same data sample and concluded
that \be \rho_{10}\approx 5.^\circ 6P^{-0.5}. \label{ro10bis} \ee

\citet{r93a,r93b} analyzed a large number of available pulse-width
data interpolated to the frequency $\sim 1$~GHz and obtained a
bimodal distribution of the opening angles. This result was later
confirmed at frequency $\sim 1.4$~GHz by \citet[][ GKS93
henceforth]{gks93} and independently by \citet{ketal94} and the
resulting $\rho-P$ relation reads \be \rho_{10}=\left\{
\begin{array}{l} 6.^\circ 3 P^{-0.5} ,
\\ 4.^\circ 9 P^{-0.5} , \end{array} \right. \label{ro10bisbis} \ee
with smaller angles preferred at shorter periods. GKS93 argued
that $4.^\circ 9$ is preferred in 80\% of pulsars with $P<0.7$~s,
and this constraint was used by GH96 (their Table~2), while ZJM03
used $4.^\circ 9$ for $P<0.7$~s and $6.^\circ 3$ for other periods
(their Table~1). We denote the former option by equation~(4a) and
the latter one by equation~(4b) in Table~2. It is worth noticing
that $\rho-P$ relations expressed by
equations~(\ref{ro10})-(\ref{ro10bisbis}) are, to some extent,
equivalent. In fact, equation~(\ref{ro10}) seems to reflect the
tendency of smaller angles being preferred at short periods in
equation~(\ref{ro10bisbis}), and equation~(\ref{ro10bis})
represents the average of both values appearing in
equation~(\ref{ro10bisbis}). Nevertheless, we use all the above
relations in our simulation procedure (section~4).

\subsection{Pulsar axes}

It is reasonable to assume that both the rotation axis and the
observer's direction are randomly oriented in space. Thus, the
probability density function for the observer angle
$\xi=\alpha+\beta$ is \be f(\xi)=\sin(\alpha+\beta). \label{f} \ee
However, the distribution of the inclination angle may depend on
many unknown factors, and the most common probability density
functions in the parent population of pulsars include the flat
function \be f(\alpha)=\frac{2}{\pi}  \label{f2} \ee for the
random distribution, the sine function \be f(\alpha)=\sin\alpha
\label{fsin} \ee for the sine-like distribution, and the cosine
function \be f(\alpha)=\cos\alpha \label{fcos} \ee for the
cosine-like distribution. One can consider some more complicated
probability density functions. For example, ZJM03 argued that the
plausible parent distribution of the inclination angle can be
described in the form \be
f(\alpha)=\frac{A}{\cosh(3.5(\alpha-0.43))}+\frac{B}{\cosh(4.0(\alpha-1.6))}
, \label{fAB} \ee where $A=0.6$ and $B=0.15$. This function has a
weak local maximum near $\alpha=\pi/2$, which is an important
feature with regard to the interpulse occurrence statistics. We
test all the above functions in our Monte Carlo simulation
procedure (section 4).

\subsection{Pulsar periods}

GH96 demonstrated that the observed distribution of periods of 516
pulsars with $4.2~{\rm s}>P>0.05$~s \citep[Pulsar
Catalog][]{tml93} can be well fitted by the gamma function \be
f(P)=G_0x^{a-1}e^{-x} \label{fP} \ee with $x=P/m$, where $m=0.3$
and $a=2.5$ ($G_0$ is the normalization constant dependent on $a$.
ZJM03 used this function to fit the period distribution in a much
larger sample of 1164 pulsar periods $4.2~{\rm s}>P>0.05$~s and
obtained $m=0.278$ and $a=2.277$. Although the parent distribution
of periods is certainly different from the observed distribution,
it is convenient to represent the former also in the form of a
general gamma function (eq.~[\ref{fP}]), with values of $m$ and
$a$ treated as free model parameters that can be derived in 2-D
grid search implemented into the Monte Carlo simulations
(section~4).

For a comparison one can also try some other trial probability
density distribution functions, like the lorentzian distribution
function \be f(P)=\frac{L_0}{1+(P-x_0)^2/a_0^2} ,\label{fPL} \ee
or even the gaussian distribution function \be
f(P)=\frac{1}{\sqrt{2\pi}\sigma_0}\exp\left[\frac{-(P-x_0)^2}{2\sigma_0}\right]
. \label{fP1} \ee Again, the values of free parameters can be
derived in 2-D grid search within the Monte Carlo simulations
(section~4).

\subsection{Detection conditions}

Following the arguments given by \citet{lm88,gh96} and
\citet{md99} we first assume that the pulsar beam is circular or
almost circular. If this is the case, then the detection condition
(see GH96 for details) is: \be \rho>|\beta| \label{ro} \ee for the
typical Main Pulse (MP) emission. However, one should also include
a rare interpulse (IP) emission, seen in a few percent of pulsars.
Two models were proposed to explain this emission occurring about
180 degrees of longitude from the centroid of the MP: the double
pole (DP-IP) model, in which MP and IP originate from the opposite
magnetic poles \citep{rl68,ml77}, and the single pole (SP-IP), in
which both MP and IP are associated with the same pole. The
detection condition for DP-IP model is \be \rho>\pi-2\alpha-\beta
\label{ropi} \ee (e.g. GH96), which favors a nearly orthogonal
rotators $(\alpha\sim\pi/2)$. Both MP and IP in pulsars that
belong to DP-IP class show relatively narrow duty cycles, contrary
to the broad profiles in the SP-IP class associated with nearly
aligned rotators $(\alpha\sim 0)$. There are two possibilities
within the SP-IP: in the first one the MP and the IP represent two
cuts through one conical beam \citep{m77,ml77}. The other
possibility is when the line-of-sight stays in the overall pulsar
beam for the entire pulsar period, so the MP and IP correspond to
cuts through two nested conical beams \citep{g83}. The detection
conditions for these two versions of SP-IP models are \be
\rho>\sqrt{2\alpha^2+\beta^2} ,\label{ro2} \ee and \be
\rho>2\alpha+\beta ,\label{ro3} \ee respectively. Both these
conditions are checked alternatively. One should mention that only
the latter condition was used by GH96, which resulted in too low
rates of SP-IP as compared to observations.

One can also consider a quite natural possibility that the pulsar
beam has a tendency to a meridional compression, with the ratio of
minor to major ellipse axes \be
R\approx\cos\frac{\alpha}{3}\sqrt{\cos(\frac{2}{3}\alpha)}
\label{R} \ee depending on the inclination angle \citep{b90,Mc93}.
Then the appropriate detection conditions (generalizations of
equations~[\ref{ro}]-[\ref{ro2}]) are: \be
|\beta|<\sqrt{R^2\rho^2+\beta^2(1-R^2)} .\label{beta} \ee for the
Main Pulse emission, \be rR>\pi-2\alpha-\beta \label{rR} \ee for
the DP-IP emission, and \be rR>2\alpha+\beta \label{rR2} \ee or
$rR>\sqrt{2\alpha^2+\beta^2}$ for the SP-IP emission, where \be
r=\sqrt{\rho^2+\beta^2\left(\frac{1}{R^2}-1\right)} \label{r} \ee
is the latitudinal beam dimension. Note that $R\leq 1$ and
$r\geq\rho$, where $\rho$ is the observationally deduced beam
radius (see GH96 for details).

\section{Observational data}

In our statistical analysis we compare the simulated distributions
with the directly ($W_{10},~P$, IP occurrences) or indirectly
$(\alpha)$ observed data by means of the Kolmogorov-Smirnov (K-S
henceforth) significance tests. Using the numerical methods given
by \citet{petal92} we compute the maximum distance ${\cal D}$
(maximum difference between cumulative distribution functions
corresponding to the observed and simulated data sets) as well as
the significance ${\cal P}$ of any non zero value of ${\cal D}$.
The value of ${\cal P}$ represents the probability that both the
simulated and the observed data sets are drawn from the same
parent distribution. We adopt an arbitrary criterion that ${\cal
P}>0.1\%$ for distributions of $P,~W_{10}$ and $\alpha$ at the
same time.

\subsection{Pulsar periods}

Using the available database of pulsar
parameters\footnote{http://www.atut.csiro.all/research/catalogue/}
we selected a sample of 1165 periods $P$ with $0.02~{\rm
s}<P<8.52$~s (Fig.~1). We rejected all recycled and binary
pulsars, since they represent different period and magnetic
inclination angle populations than typical pulsars (the recycled
pulsar B1933+16 with $P=0.089$~s was also excluded). This sample
of the observed pulsar periods can be fitted by the gamma function
expressed by equation~(\ref{fP}) with $m=0.28$ and $a=2.28$ (see
upper panel in Fig.~1).

\subsection{Pulse-widths}

We use the database of 238 pulse-width measurements $W_{10}$ taken
at the frequency of about 1.4 GHz (upper panel in Fig.~2),
compiled by GH96 (their Table~1 and Fig.~1)\footnote{We excluded 4
recycled and binary pulsars from the database of GH96.}. These
measurements were carefully selected from the available databases
to satisfy all criterions imposed by the symmetry of
equation~(\ref{w10}), (see comments below this equation and GH96
for more details). ZMJ03 demonstrated that another available
database (containing 265 pulsars) of \citet{gl98} is equivalent to
that of GH96 in the sense that the values of $W_{10}$ are roughly
the same in both these databases.

\subsection{Inclination angles}

While the values of $W_{10}$ and $P$ as well as the interpulse
occurrences are direct observational quantities, the values of
inclination angles $\alpha$ can only be indirectly observed by
means of the polarization measurements
\citep{lm88,r90,r93a,r93b,g94}. Following the arguments of ZJM03
we used the database of the inclination angles compiled by
\citet{r93a,r93b}. This database contains 149 measurements of the
magnetic inclination angles, whose distribution is presented in
the upper panel of Fig.~3.

\subsection{Interpulse emission}

The fraction of interpulses in the observed sample of pulsars
provides useful information about pulsar geometry (see GH96).
However, the statistical studies of this phenomenon are difficult
since the ratio of amplitudes of the interpulse to the main-pulse
is often about 1\% and varies with frequency \citep{hf86}. The
only representative sample of interpulses can be found in Table 6
of the Catalog of 558 pulsars \citep{tml93}. There are 22 pulsars
with interpulses $(\sim 4\%)$ in this sample, with only 3 certain
cases (B0826-34, B0950+08, B1929+10) corresponding to a single
magnetic pole \citep[][ their Fig.~8]{lm88}. Most likely PSR
B1848+04 with very a broad main-pulse also belongs to this
category. This corresponds to occurrence rates of about 3.4\% and
0.5\% for DP-IP and SP-IP, respectively. These rates were adopted
as model values by GH96. However, to be consistent with our
selection of pulsar periods (section~3.2) one should exclude
millisecond and other recycled pulsars from both the total pulsar
sample and from the sample of pulsars with interpulses. When this
is done then the corresponding rates of occurrence are 2\% and
0.8\% for DP-IP and SP-IP, respectively. For a comparison, in the
published sample of 420 pulsars found recently in the Parkes
multibeam pulsar survey \citep{metal01,metal02,ketal04} one can
identify clearly 4 cases of DP-IP and 1 case of SP-IP, and a few
low intensity candidates. This gives lower limits of about 1\% and
0.24\%, respectively. It is difficult to estimate the actual rates
of IP occurrences until the sensitive search for interpulses in
the newly discovered pulsars is made. We assume that the plausible
distributions of periods $P$ and inclination angles $\alpha$ in
the parent pulsar population should be able to explain about 2\%
of DP-IP and slightly above 0.5\% of SP-IP in the normal pulsar
population (excluding millisecond and other recycled pulsars).

\section{Monte Carlo simulations}

We performed the Monte Carlo simulations of the pulse widths
$W_{10}$, pulsar periods $P$ and inclination angles $\alpha$, in
an attempt to reproduce the observed distributions of these
quantities. We have used the random number generator given in
\citet{petal92}. Our simulation procedure can be described as a
number of subsequent steps:
\begin{description}
\item (1) Generate the inclination angle $\alpha$ as a random
deviate with the parent probability density function $f(\alpha)$
corresponding to equations (\ref{f2}), (\ref{fsin}), (\ref{fcos})
and (\ref{fAB}), respectively.
\item (2) Generate the observer angle $\xi=\alpha+\beta$ as a
random deviate with the parent probability density function
$f(\xi)=\sin\xi$ (eq.~[\ref{f}]), and calculate the impact angle
$\beta=\xi-\alpha$.
\item (3) Generate the pulsar period $P$ as a random deviate with
the parent probability density function $f(P)$ corresponding to
equations~(\ref{fP}), (\ref{fPL}) and (\ref{fP1}), respectively.
Record the free model parameters of each function used.
\item (4) For a given value of $P$ calculate the opening angle
$\rho_{10}$ corresponding to equations~(\ref{ro10}),
(\ref{ro10bis}) and (\ref{ro10bisbis}), respectively.
\item (5) Check the detection condition corresponding to
equations~(\ref{ro}), (\ref{ropi}) and (\ref{ro2}) for
$N_{tot}=50000$ simulated pulsars for each combination of $\rho-P$
relations (eqs.~[\ref{ro10}]-[\ref{ro10bisbis}]) and distribution
functions $f(P)$ (eqs.~[\ref{fP}]-[\ref{fP1}]) and $f(\alpha)$
(eqs.~[\ref{f2}]-[\ref{fAB}]). Record the number of observed
pulsars $N_{obs}$. Calculate the beaming fraction
$f=N_{obs}/N_{tot}$, as well as the rates of detected interpules
from two magnetic poles (DP-IP) and a single magnetic pole
(SP-IP).
\item (6) For each set of parameters $\alpha,~\beta$ and
$\rho_{10}(P)$ corresponding to the observable pulsar calculate
the pulse width $W_{10}$ according to equation~(\ref{w10}), and
record the relevant information $(W_{10},P,\alpha)$.
\item (7) Judge the statistical significance using the
K-S tests for the simulated and observed distributions of pulse
width $W_{10}$, period $P$ and inclination angle $\alpha$.
\end{description}

\subsection{Reproduction of previous results of GH96 and ZMJ03}

To make sure that our simulation software works properly, we begun
with a reproduction of results given in the previous statistical
works of GH96 and ZJM03. As expected, we managed to reproduce the
K-S statistics for $W_{10}$, beaming fraction $f$ and the rates of
observed interpulses given in Table~2 of GH96 (to save space we do
not reproduce this table here). However, the K-S test for pulsar
periods (not included in GH96) resulted in very low probabilities
(below $10^{-6}$), confirming the suggestion of TM98 that the
distribution of periods in the parent pulsar population is
significantly different from that of the observed distribution. As
we argue later on in this paper, the plausible parent distribution
of periods can be expressed in the form of a gamma function
(eq.~[\ref{fP}]) with $m=0.34\pm 0.02$ and $a=2.52\pm 0.04$, in
contrast to $m=0.3$ and $a=2.5$ obtained for the observed sample
of 516 pulsar periods by GH96, or $m=0.28$ and $a=2.28$ obtained
for 1165 periods used in this paper (see also ZJM03).

Twelve entries in Table~1 correspond to cases A1-A4, B1-B4 and
C1-C4 from Table~1 in ZJM03 (we excluded cases D1-D4 since for
$s=5.8/8.8=0.65$ they are equivalent to cases A1-A4). As one can
see, we reproduced quite well the results of K-S tests for $P$,
$W_{10}$ and $\alpha$, as well as the values of beaming fraction
$f$ (our period sample has few more pulsars below 0.05~s and above
4.2~s as compared to the one used by ZJM03). We added the rates of
the interpulse occurrence and concluded that the pure $\cos\alpha$
distribution of the parent inclination angles (A1, B1 and C1)
generates much too few DP-IP (about 0.3\% as compared with about
2\% observed). Thus, although the pure $\cos\alpha$ distribution
gives very good results of K-S tests (consistent with the results
of TM98 and ZJM03), it should be rejected on the grounds of
unacceptable interpulse statistics. We found out that the modified
cosine distribution of ZJM03 (see eq.~[9]) not only gives the
plausible results of K-S test, but also reproduces the rate of
occurrence of both DP-IP $(\sim 2\%)$ and SP-IP $(0.5\%-0.7\%)$.

\subsection{Other new results}

We have examined 19200 combinations of distribution functions of
pulsar periods (eqs. [\ref{fP}], [\ref{fPL}] and [\ref{fP1}] with
200 combinations of parameters in each case), opening angles
(eqs.~[\ref{ro10}], [\ref{ro10bis}], [4a] and [4b]), inclination
angles (eqs.~[\ref{f2}], [\ref{fsin}], [\ref{fcos}] and
[\ref{fAB}]) as well as two options of the beam shape: circular
(eqs.~[\ref{ropi}]-[\ref{ro3}]) and elliptical
(eqs.~[\ref{R}]-[\ref{r}]). We recorded only those cases in which
the following conditions were simultaneously satisfied: the
probabilities ${\cal P}$ that the observed and simulated
distributions of $P,~W_{10}$ and $\alpha$ exceeded 0.1\%, as well
as the rates of occurrence of DP-IP and SP-IP exceeded 2\% and
0.5\%, respectively. With the adopted step of 0.02 in the
parameters of gamma, lorentzian and gaussian functions
(eq.~[\ref{fP}]-[\ref{fP1}]), this resulted in 15 records. Table~2
presents 5 representative cases with the highest rates of SP-IP
occurrence. We believe that case (1) is the most plausible one and
Figs.~1-3 present a visual comparison of the observed and
simulated distribution corresponding to this case. Below we
discuss some aspects of our analysis, which are not reflected in
Table~2.

\subsubsubsection{Pulsar period}

Both gamma (eq.~[\ref{fP}]) and lorentzian (eq.~[\ref{fPL}])
functions with parameter values $m=0.34\pm 0.02$, $a=2.52\pm 0.04$
and $x_0=0.62\pm 0.02$, $a_0=0.4\pm 0.02$ respectively, are
plausible parent density distribution functions. However, the
gamma function seems much better suited to reflect the skew
character of the pulsar period distribution. The gaussian function
(eq.~[\ref{fP1}]) is rather unlikely. In the best case
corresponding to $x_0=0.73$ and $\sigma_0=0.36$, the significance
${\cal P}$ is only about 0.03\%

\subsubsubsection{Inclination angle}

ZJM03 argued, ignoring the issue of the interpulse emission, that
the parent distribution of the magnetic inclination angles can be
expressed by the cosine function or modified cosine function
represented by equation~(\ref{fAB}) in this paper. Including the
analysis of interpulse statistics we confirm the latter, but we
refute the cosine function, since it generates less than 0.3\% of
DP-IP (as compared with about 2\% observed). This is a strong
conclusion since better statistics of interpulse occurrences
cannot change it in the future. As for other trial distribution
functions, the sine-like function (eq.~[\ref{fsin}]) generates too
many DP-IP $(\sim 5\%)$ and far too few SP-IP $(\sim 0.02\%)$,
while the interpulse rates generated by the flat distribution
function (eq.~[\ref{f2}]) are roughly comparable with
observations, especially when associated with the meridionally
compressed beam (eq.~[\ref{R}]), but the results of K-S test for
the magnetic inclination angles are not promising $({\cal
P}<10^{-5})$. Therefore, the modified cosine function
(eq.~[\ref{fAB}]) is the only plausible density distribution
function that satisfies all constraints. We have checked whether
one can improve the statistical results by changing values of
parameters $A$ and $B$ in equation~(\ref{fAB}). It appeared that
without violating the basic condition that ${\cal P}>0.1\%$ for
all considered quantities ($P,~W_{10}$ and $\alpha$), one can only
increase the interpulse rates by a small fraction (e.g. from
0.74\% to 0.77\% for SP-IP) and thus indeed $A\backsimeq 0.6$ and
$B\backsimeq 0.15$ as suggested by ZJM03.

\subsubsubsection{Beaming fraction}

For the 15 cases satisfying all constraints adopted in our
analysis, the beaming fraction $f$ defined as the number of
detected pulsars divided by 50000 detection attempts is $0.124\pm
0.004$. This is consistent with the probabilities of observing a
normal radio pulsar obtained by both TM98 and ZJM03.

\section{Conclusions and Discussion}

In this paper we performed statistical studies using the Monte
Carlo simulations of the possible parent distributions of pulsar
periods $P$ and magnetic inclinations angles $\alpha$. We
generated synthetic distributions of the pulse-widths $W_{10}$, as
well as the interpulse occurrences, and confronted them with the
observational data by means of the Kolmogorov-Smirnov significance
tests. We found out that the observed distributions of pulsar
periods, pulse widths and inclination angles are relatively easy
to reproduce with a variety of trial density distribution
functions. However, when we use the criterion that the K-S
significance probabilities ${\cal P}$ for $P,~W_{10}$ and $\alpha$
are higher than 0.1\% and that the generated interpulse rates
agree with the observed rates, then we were left with just a few
possibilities presented in Table~2. Our results can be summarized
as follows:
\begin{description}
\item (1) The distribution of the magnetic inclination angles in
the parent pulsar population is a complicated function represented
by equation~(\ref{fAB}), which has a local maximum around
$\alpha=25^\circ$ and another weaker one around $\alpha=90^\circ$.
This function reproduces the observed distribution of the observed
inclination angles (Fig.~3), as well as generates the rates of
occurrence of interpulses at the observed levels.
\item (2) The parent distribution of periods in the normal pulsar
population (excluding millisecond and other recycled pulsars) can
be described by the gamma function (eq.~[\ref{fP}]) with
parameters $m=0.34\pm 0.02$ and $a=2.52\pm 0.04$. The actual
values of these parameters differ slightly depending on the
$\rho$-P relation.
\item (3) There is no evidence that the shape of the pulsar beam
deviates significantly from a circular crossection. It is likely
that the internal structure of a typical pulsar beam consist of
two co-axial cones (eq.~[\ref{ro10bisbis}])
\item (4) The beaming fraction $f$, that is the fraction of observable
pulsars or the probability of observing a normal pulsar, is
$0.124\pm 0.004$.
\end{description}

In general, our results are consistent with those of ZJM03.
However, we added the constrains related to the interpulse
analysis and found out that this aspect of our analysis is the
most restrictive one. We were able to reject the pure cosine
distribution of the inclination angles, but we confirmed their
modified cosine function which has a local maximum around
$\alpha=25^\circ$ and another weaker one near $\alpha=90^\circ$.
This means that the evolution of the magnetic inclination angles
in pulsars cannot be described by any simple law (alignment,
counteralignment, etc).

Although we improved the analysis of ZJM03 by adding the
interpulse statistics, we are still missing an analysis of a
possible effect of the intrinsic luminosity of radio pulsars on
our results. This problem is, however, very difficult and
complicated and we will postpone a full treatment to the
subsequent paper. The proper approach would be to compare the
synthetic radio luminosity with the minimum detectable flux
achieved in a given pulsar survey, and thus it can be applied only
to a uniform data sets of pulsars detected in single survey. Our
data do not have such a degree of uniformity. However, most
surveys were less sensitive to long-period pulsars, as it follows
from the nature of applied Fourier-transform method. Since the
interpulse emission (which appears to be the most restrictive
constraint in our analysis) occur mainly at shorter periods, then
a possible underrepresentation of pulsars with longer periods
should not affect significantly our general results.

\begin{acknowledgements}
This work is partially supported by the Grant 1 P03D 029 26 of the
Polish State Committee for Scientific Research. We thank E. Gil
and U. Maciejewska for technical help. We also thank and anonymous
referee for very constructive and helpful criticism.
\end{acknowledgements}

{}

\begin{deluxetable}{cp{0pt}cp{0pt}ccp{0pt}ccp{0pt}ccp{0pt}ccp{0pt}cp{0pt}cc}
\tabletypesize{\scriptsize} \rotate \tablecaption{Results of Monte
Carlo simulations - reproduction of Table~1 in ZJM03 \label{tbl1}}
\tablewidth{0pt} \tablehead{$\rho-P$ & & Inclination &&
\multicolumn{2}{c}{Pulsar} &&
\multicolumn{8}{c}{Kolmogorov-Smirnov} &&
Beaming  && \multicolumn{2}{c}{Interpulse}\\
relation\tablenotemark{\dag} & & angle & &
\multicolumn{2}{c}{period} && \multicolumn{8}{c}{
statistics\tablenotemark{\dag\dag\dag}}
 && fraction && \multicolumn{2}{c}{rates}\\
\cline{1-1}  \cline{3-3} \cline{5-6} \cline{8-15} \cline{17-17}
\cline{19-20} $\rho_{10} (P)$ & & $f(\alpha)$ & &
\multicolumn{2}{c}{$f(P)$\tablenotemark{\dag\dag}} & &
\multicolumn{2}{c}{$P$} && \multicolumn{2}{c}{$W_{10}$} &&
\multicolumn{2}{c}{$\alpha$} & & &\\ \cline{5-6} \cline{8-9}
\cline{11-12} \cline{14-15} && && $m$ & $a$ && ${\cal D}$ & ${\cal
P}$ && ${\cal D}$ & ${\cal P}$ && ${\cal D}$ & ${\cal P}$ && f &&
DP-IP & SP-IP} \startdata \vspace{-13pt}\\ \hline A. & &
\multicolumn{1}{l}{1. $\cos\alpha$} && 0.38 & 2.21 && 0.049 &
0.017
&& 1.068 & 0.236 && 0.139 & 0.007 && 0.119 && 0.25\% & 0.62\%  \\
\hline A. & & \multicolumn{1}{l}{2. eq.(9)} && 0.38 & 2.204 &&
0.046
& 0.029 && 0.108 & 0.009 && 0.102 & 0.093 && 0.123 && 2.24\% & 0.36\% \\
\hline A. && \multicolumn{1}{l}{3. $2/\pi$} && 0.278 & 2.277 &&
0.151 & $10^{-18}$ && 0.048 & 0.666 && 0.354 & $10^{-13}$ & &0.162
&& 3.58\% & 0.39\% \\ \hline A. && \multicolumn{1}{l}{4. $2/\pi$}
& & 0.411 & 2.101 && 0.046 & 0.028 && 0.105 & 0.012 &&
0.360 & $10^{-14}$ && 0.147 && 3.37\% & 0.41\% \\
\hline\hline B. & & \multicolumn{1}{l}{1. $\cos\alpha$} & & 0.39 &
2.339 && 0.054 & 0.007 && 0.121 & 0.002 && 0.140 & 0.006 && 0.121&
& 0.28\% & 1.01\%
\\ \hline B. && \multicolumn{1}{l}{2. eq.(9)} & & 0.4 & 2.29 && 0.051 &
0.011 && 0.157 & $10^{-5}$ && 0.1 & 0.110 && 0.126 && 2.58\% & 0.69\% \\
\hline B. && \multicolumn{1}{l}{3. $2/\pi$} & & 0.278 & 2.277 &&
0.184 & 0.007 && 0.148 & $10^{-4}$ && 0.347 & $10^{-13}$ && 0.179
&& 4.71\% & 0.75\% \\ \hline B. && \multicolumn{1}{l}{4. $2/\pi$}
& & 0.4 & 2.255 && 0.049 & 0.017 &
&0.101 & 0.018 && 0.354 & $10^{-14}$ && 0.154 && 3.72\% & 0.49\% \\
\hline\hline C. && \multicolumn{1}{l}{1. $\cos\alpha$} & & 0.38 &
2.27 && 0.054 & 0.006 && 0.086 & 0.066 && 0.137 & 0.008 && 0.122&
& 0.29\% & 0.90\%
\\ \hline C. && \multicolumn{1}{l}{2. eq.(9)} & & 0.37 & 2.27 && 0.057 &
0.009 && 0.137 & 0.019 && 0.1 & 0.107 && 0.128 && 2.54\% & 0.51\% \\
\hline C. && \multicolumn{1}{l}{3. $2/\pi$} & & 0.278 & 2.277 &&
0.152 & $10^{-18}$ && 0.07 & 0.206 && 0.352 & $10^{-14}$ && 0.168
&& 4.12\% & 0.57\% \\ \hline C. && \multicolumn{1}{l}{4. $2/\pi$}
& & 0.378 & 2.262 && 0.047 & 0.023
&&0.085 & 0.067 && 0.355 & $10^{-14}$ && 0.153 && 3.57\% & 0.43\% \\
\hline \enddata \tablenotetext{\dag}{$\rho-P$ relation (beam
radius): A -- $\rho_{10}=5.^\circ 8 P^{-1/3}$ (eq.~[2]), B --
$\rho_{10}=5.^\circ 6 P^{-1/2}$ (eq.~[3]),\\ \hspace*{4.3cm} C --
$\rho_{10}=4.^\circ 9 P^{-1/2}$ for $P<0.7$~s and
$\rho_{10}=6.^\circ 3 P^{-1/2}$ for others (eq.~[4b])}
\tablenotetext{\dag\dag}{gamma function
$f(P)=G_0(P/m)^{a-1}e^{-(P/m)}$ (eq.~[10])}
\tablenotetext{\dag\dag\dag}{K-S tests: ${\cal D}$ -- maximum
difference between cumulative distribution functions corresponding
to the observed and the simulated data samples,\\
\hspace*{2cm} ${\cal P}$ -- probability that both the observed and
the simulated data sets are drawn from the same parent
distribution.}
\end{deluxetable}

\setcounter{table}{1}
\begin{deluxetable}
{ccp{0pt}cp{0pt}ccp{0pt}ccp{0pt}ccp{0pt}ccp{0pt}cp{0pt}cc}
\tabletypesize{\scriptsize} \rotate \tablecaption{ Results of
Monte Carlo simulations \label{tb21}} \tablewidth{0pt}
\tablehead{& $\rho-P$ & & Inclination &&
\multicolumn{2}{c}{Pulsar} &&
\multicolumn{8}{c}{Kolmogorov-Smirnov} &&
Beaming  && \multicolumn{2}{c}{Interpulse}\\
$No$ & relation & & angle & & \multicolumn{2}{c}{period} &&
\multicolumn{8}{c}{ statistics\tablenotemark{\dag}}
 && fraction && \multicolumn{2}{c}{rates}\\
\cline{2-2}  \cline{4-4} \cline{6-7} \cline{9-16} \cline{18-18}
\cline{20-21} & $\rho_{10} (P)$ & & $f(\alpha)$ & &
\multicolumn{2}{c}{$f(P)$} & & \multicolumn{2}{c}{$P$} &&
\multicolumn{2}{c}{$W_{10}$} && \multicolumn{2}{c}{$\alpha$} & &
&\\ \cline{6-7} \cline{9-10} \cline{12-13} \cline{15-16} & && && &
&& ${\cal D}$ & ${\cal P}$ && ${\cal D}$ & ${\cal P}$ && ${\cal
D}$ & ${\cal P}$ && f && DP-IP & SP-IP} \startdata \vspace{-13pt}\\
\hline 1. & eq.~(4a) & & \multicolumn{1}{l}{eq.~(9)} &&
\multicolumn{2}{c}{eq.~(10)} && &
&& & && & && && & \\
& & & && $m=0.34$ & $a=2.52$ && 0.054 & 0.006 &
& 0.124 & 0.002 && 0.093 & 0.156 && 0.128 && 2.07\% & 0.74\% \\
\hline 2. & eq.~(3) & & \multicolumn{1}{l}{eq.~(9)} &&
\multicolumn{2}{c}{eq.~(10)} && &
&& & && & && && & \\
& & & && $m=0.36$ & $a=2.6$ && 0.055 & 0.006 &
& 0.127 & 0.001 && 0.101 & 0.103 && 0.121 && 2.55\% & 0.64\% \\
\hline 3. & eq.~(4b) & & \multicolumn{1}{l}{eq.~(9)} &&
\multicolumn{2}{c}{eq.~(10)} && &
&& & && & && && & \\
& & & && $m=0.32$ & $a=2.6$ && 0.054 & 0.006 &
& 0.117 & 0.004 && 0.098 & 0.120 && 0.125 && 2.44\% & 0.53\% \\
\hline 4. & eq.~(2) & & \multicolumn{1}{l}{eq.~(9)} &&
\multicolumn{2}{c}{eq.~(11)} && &
&& & && & && && & \\
& & & && $x_0=0.6$ & $a_0=0.42$ && 0.053 & 0.008 &
& 0.090 & 0.049 && 0.107 & 0.072 && 0.123 && 2.38\% & 0.57\% \\
\hline 5. & eq.~(4b) & & \multicolumn{1}{l}{eq.~(9)} &&
\multicolumn{2}{c}{eq.~(11)} && &
&& & && & && && & \\
& & & && $x_0=0.64$ & $a_0=0.38$ && 0.058 & 0.003 &
& 0.115 & 0.004 && 0.109 & 0.061 && 0.126 && 2.25\% & 0.51\% \\
\hline \enddata \tablenotetext{\dag}{see Table~1 for explanation}
\end{deluxetable}

\begin{table}[t]
\begin{center}
\setcounter{table}{2} \caption{Interpulse emission in normal
pulsars (after Table~6 in Taylor et al.~1993)}
\begin{tabular}[t]{cccclcc}
\hline
No. & PSR & $P$ (sec)& & Fractional & Phase & How many \\
    &     &          & & Amplitude & separation $(^\circ)$ & poles? \\
    \hline
1. & B0531$+$21 & 0.033 & & 0.6 & 145 & DP \\
2. & B0823$+$26 & 0.530 & & 0.005 & 180 & DP\\
3. & B0826$-$34 & 1.848 & & 0.1 & 180 & SP\\
4. & B0906$-$49 & 0.106 & & 0.24 & 176 & DP\\
5. & B0950$+$08 & 0.253 & & 0.012 & 210 & SP\\
6. & B1055$-$52 & 0.197 & & 0.5 & 205 & DP\\
7. & B1259$-$63 & 0.047 & & 0.75 & 145 & DP\\
8. & B1702$-$19 & 0.298 & & 0.15 & 180 & DP\\
9. & B1719$-$37 & 0.236 & & 0.15 & 180 & DP\\
10. & B1736$-$29 & 0.322 & & 0.4 & 180 & DP\\
11. & B1822$-$09 & 0.768 & & 0.05 & 185 & DP\\
12. & B1848$+$04 & 0.285 & & 0.2 & 200 & SP\\
13. & B1929$+$10 & 0.226 & &0.018 & 170 & SP\\
14. & B1944$+$17 & 0.441 & &0.005 & 175 & DP\\
\hline
\end{tabular}
\end{center}
\end{table}

\begin{figure}[!t]
\begin{center}
\includegraphics[width=12cm,height=20cm]{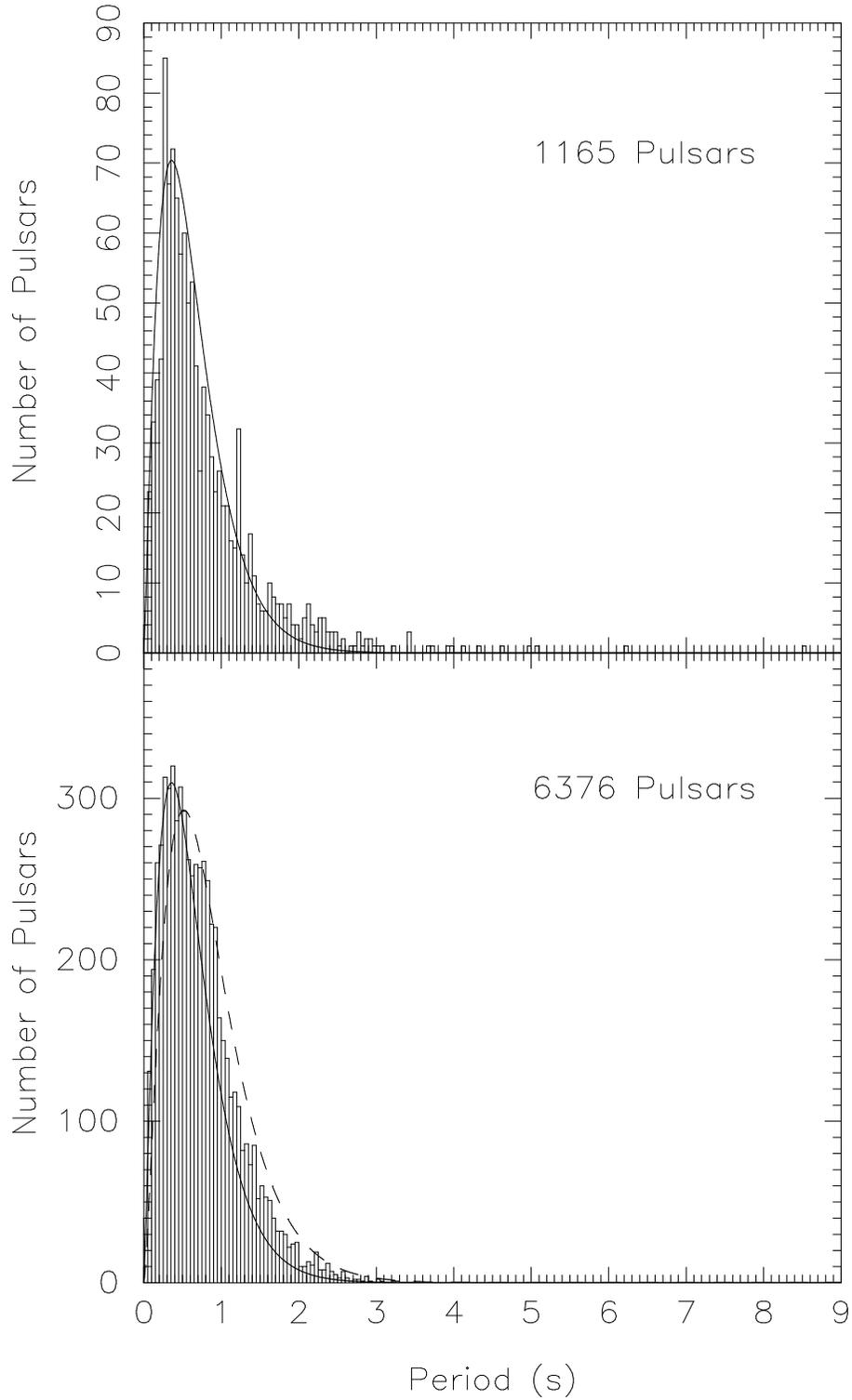}
\caption{Observed distribution of 1165 pulsar periods with
$0.02~{\rm s}<P<8.52$~s (upper panel) and simulated distribution
corresponding to case (1) in Table~2 (lower panel). The solid
lines represent the analytical fit to the observed distribution in
the form of a gamma function with $m=0.28$ and $a=2.28$, while the
dashed line represents the parent density distribution gamma
function with $m=0.34$ and $a=2.52$.}
\end{center}
\end{figure}

\begin{figure}[!t]
\begin{center}
\includegraphics[width=12cm,height=20cm]{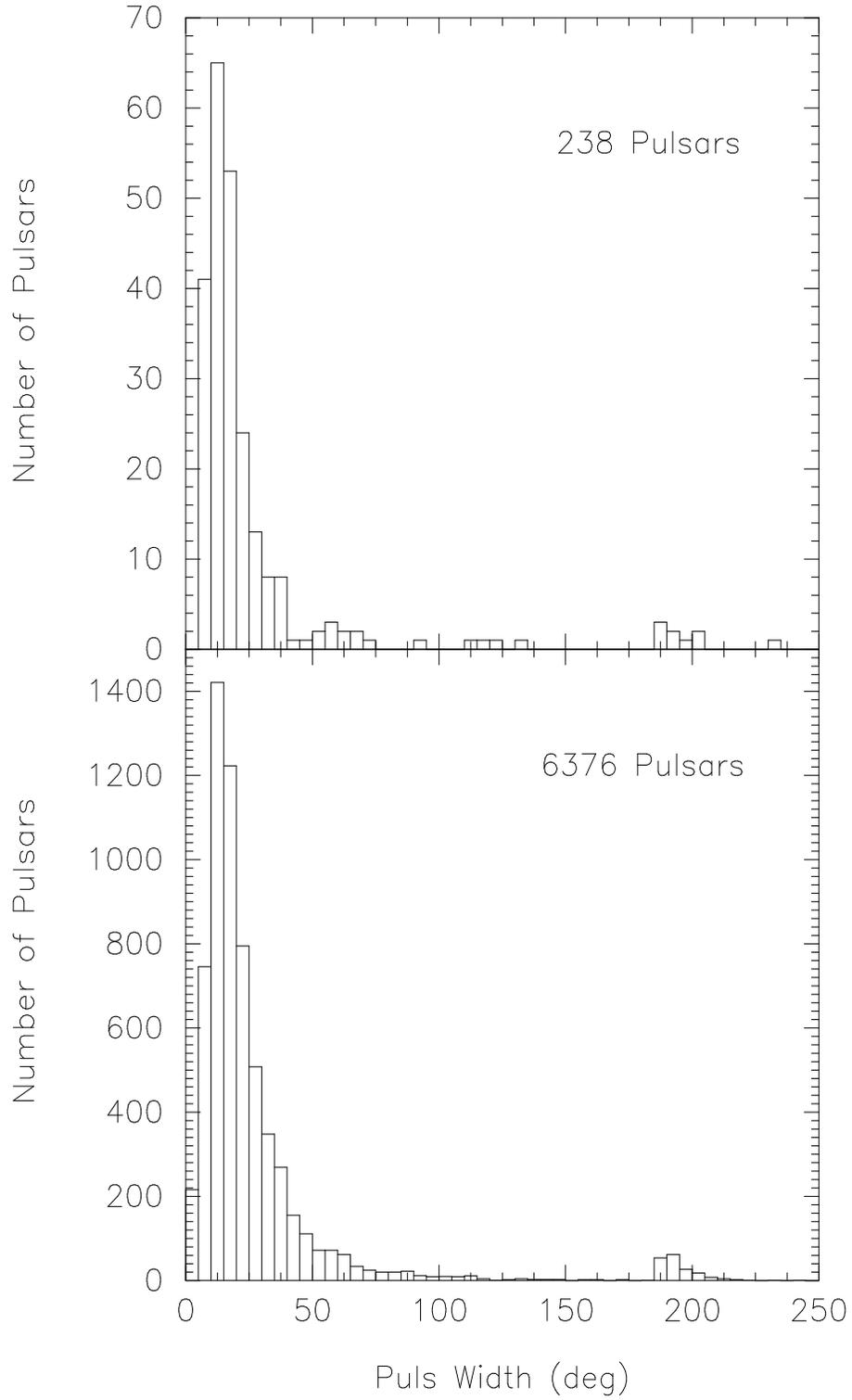}
\caption{Observed distribution of 238 pulse widths $W_{10}$ taken
from GH96 (upper panel) and simulated distribution corresponding
to case (1) in Table~2 (lower panel).}
\end{center}
\end{figure}

\begin{figure}[!t]
\begin{center}
\includegraphics[width=12cm,height=20cm]{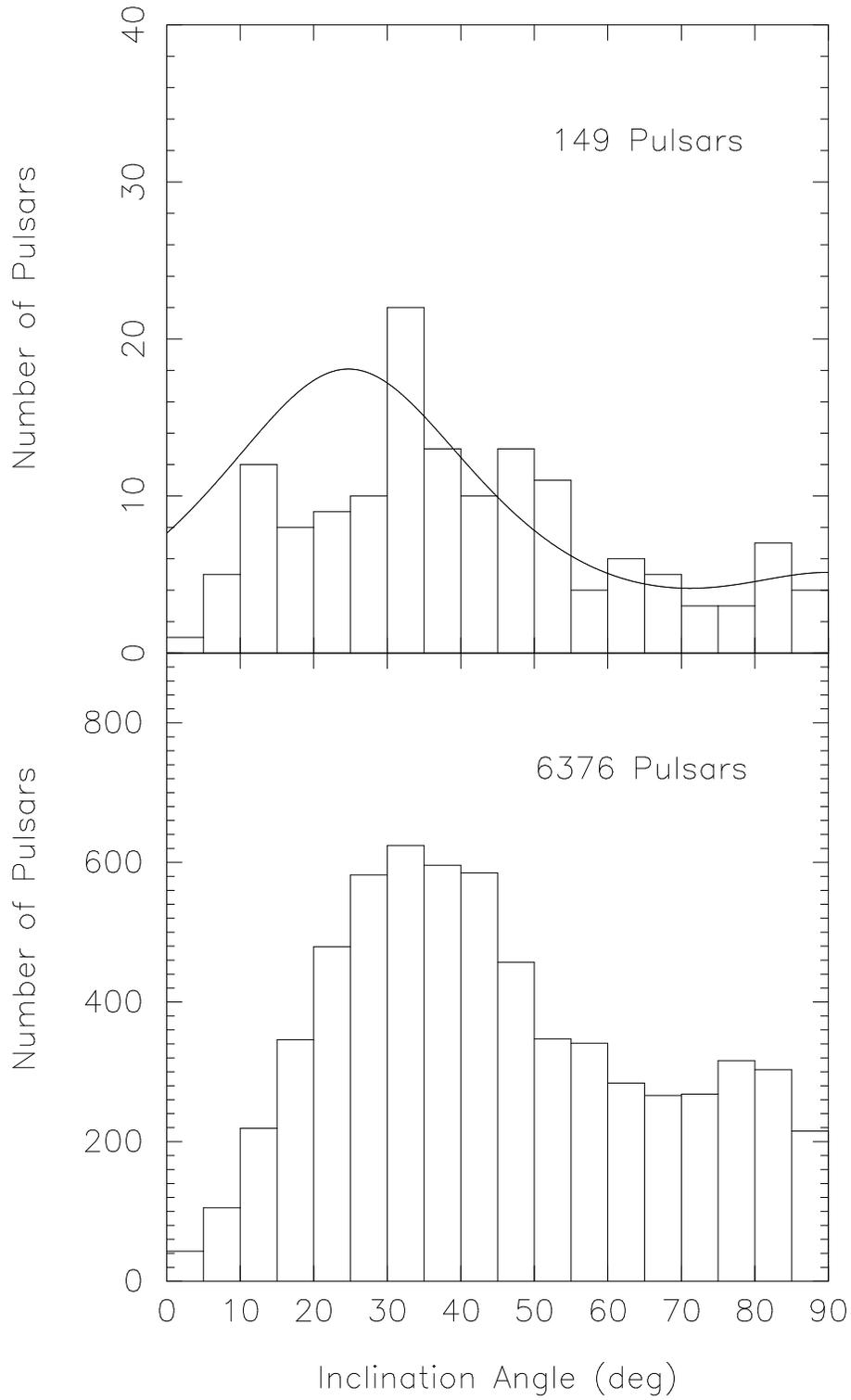}
\caption{Observed distribution of 149 magnetic inclination angles
$\alpha$ taken from \citet{r93b} (upper panel) and simulated
distribution corresponding to case (1) in Table~2 (lower panel).
The solid line represents the parent density distribution function
expressed by equation~(\ref{fAB}).}
\end{center}
\end{figure}

\end{document}